\documentclass[12pt,letterpaper]{article}
\usepackage{citesort}
\usepackage[xdvi]{graphicx}
\usepackage{sccs}  
\usepackage{amsmath,amssymb}
\usepackage{times}

\newcommand{\bfr}{\ensuremath{\mathbf r}}
\newcommand{\bfk}{\ensuremath{\mathbf k}}

\newcommand{\rh}{\tilde{\rho}_T}
\newcommand{\calp}{\ensuremath{\mathcal P}}

\newcommand{\sidebyside}[2]{
  \parbox{.5\linewidth}{\centerline{#1}\centerline{(a)}}
  \parbox{.5\linewidth}{\centerline{#2}\centerline{(b)}}
  \vspace{0.5cm}}
\newcommand{\sidebysidenl}[2]{
  \parbox{.5\linewidth}{\centerline{#1}}
  \parbox{.5\linewidth}{\centerline{#2}}}

\begin{document}
\title{Path Integral Monte Carlo Simulations for Fermion Systems: \\
       Pairing in the Electron-Hole Plasma.}
\author{J.~Shumway and D.~M.~Ceperley}
\address{National Center for Supercomputing Applications and
         Department of Physics,\\
         University of Illinois Urbana-Champaign, Urbana, Illinois, 61801}
\maketitle 
\begin{abstract}
We review the path integral method wherein quantum systems are
mapped with Feynman's path integrals onto a classical system of
``ring-polymers'' and then simulated with the Monte Carlo
technique. Bose or Fermi statistics correspond to possible
``cross-linking'' of polymers. As proposed by Feynman,
superfluidity and Bose condensation result from macroscopic
exchange of bosons.  To map fermions onto a positive probability
distribution, one must restrict the paths to lie in regions where
the fermion density matrix is positive. We discuss a recent
application to the two-component electron-hole plasma. At low
temperature excitons and bi-excitons form. We have used nodal
surfaces incorporating paired fermions and see evidence of a Bose
condensation in the energy, specific heat and superfluid density.
In the restricted path integral picture, pairing appears as
intertwined electron-hole paths. Bose condensation occurs when
these intertwined paths wind around the periodic boundaries.
\end{abstract}

\section{INTRODUCTION}

As first shown by Feynman~\cite{feynman53}, the thermodynamics of a
quantum many-body system can be investigated using
classical-statistical methods on polymer-like systems. To do this,
one writes the density matrix of a many-body system at a temperature
$k_B T = \beta^{-1}$ as an integral over all paths $\{ R_t \}$:
\begin{equation}
\rho (R_0, R_{\beta} ; \beta) = \frac{1}{N!} \sum_{\calp} (\pm
1)^{\calp} \oint_{\calp R_0 \rightarrow R_{\beta}}
 dR_t \exp ( - S[R_t ]).
\end{equation}
The path $R(t)$ begins at $\calp R_0$ and ends at $
R_{\beta}$, and \calp~ is a permutation of particle labels. For
$N$ particles, the path is in $3N$ dimensional space: $R_t =
(\bfr_{1t} , \bfr_{2t} \ldots \bfr_{Nt} )$. The upper sign is to
be used for bosons and the lower sign for fermions. For
nonrelativistic particles interacting with a potential $V(R)$, the
{\it action} of the path, $ S[R_t ]$,  is given by:
\begin{equation}
S[R_t ] = \int_0^{\beta} dt \left[ \frac{m}{2} \left|
\frac{dR_t}{\hbar dt} \right|^2 + V(R_t) \right].
\end{equation}
Thermodynamic properties, such as the energy, are related to the
diagonal part of the density matrix, so that the path returns to
its starting place or to a permutation \calp~ of its starting
place after a ``time'' $\beta$.

Since the imaginary-time {\it action } $ S[R_t ]$ is a real
function of the path, for boltzmannons or bosons the integrand is
nonnegative, and can be interpreted as a probability of an
equivalent classical system and the action as the classical
potential energy of a ``polymer.'' To perform Monte Carlo
calculations of the integrand, one  makes imaginary time discrete,
so that one has a finite number of time slices and thereby a
classical system of $N$ particles in $M$ time slices. If the path
integral is performed by a simulation method, such as a
generalization of Metropolis Monte Carlo or with molecular
dynamics, one can obtain essentially exact results for problems
such as the properties of liquid $^4$He at temperatures near the
superfluid phase transition~\cite{ceperley95}.

In addition to sampling the path, the permutation is also sampled.
This is equivalent to allowing the ring polymers to connect in
different ways. This macroscopic ``percolation'' of the polymers
is directly related to superfluidity~\cite{feynman53}. Superfluid
behavior can occur at low temperature when the probability of
exchange cycles on the order of the system size is nonnegligible.
For details see Ref.~\cite{ceperley95}.

However, the straightforward application of those techniques to
Fermi systems means that odd permutations subtract from the
integrand.  The calculation of any physical operator by direct
fermion PIMC is very inefficient because of the cancellation of
positive and negative permutations. This is the ``fermion sign
problem.'' Path integral methods as rigorous and successful as
those for boson systems are not yet known for fermion systems in
spite of the activities of many scientists throughout the last
four decades.

\section{RESTRICTED PATH INTEGRALS}

For the diagonal density matrix we can arrange things so that we
only get positive contributions by restricting the paths. We now
sketch the derivation of the restricted path identity: that the
nodes of the exact fermion many-body density matrix determine the
rule by which one can take only paths with the same sign. The
nodes of the fermion density matrix $\rho_F (R, R_* ; t)$ carve up
space-time into a finite number of nodal cells defined as follows:
call a {\it node-avoiding path}, a continuous path $R_t$ for $0 <
t \leq \beta$ for which $\rho_F (R_t, R_* ; t) \ne 0$ for all $0 <
t < \beta$. Two points are in the same nodal cell if they are
connected by a node-avoiding path (with respect to the same fixed
$R_*$). The collection of all space time points $(R,t)$ connected
by some node-avoiding path make up a nodal cell. Then we can solve
the Bloch equation inside each nodal cell separately by specifying
the initial conditions at $t=0$ and zero boundary conditions on
the surface of the nodal cell. To enforce the zero boundary
conditions, we insert an infinite repulsive potential precisely at
the nodal surfaces which eliminates the contribution of any walks
which hit or cross the node. Thus:
\begin{equation}
\label{RPI} \rho_F (R_{\beta} ,R_*;\beta)  =  \int
dR_0 \rho_F (R_0, R_* ; 0)  \oint_{R_0 \rightarrow R_{\beta}
\in \Upsilon (R_*)} dR_t e^{-S[R_t] }
\end{equation}
where the subscript means that we restrict the path integration to
paths starting at $R_0$,  ending at $R_{\beta}$ and are
node-avoiding. The weight of the walk is $\rho_F (R_0, R_* ; 0)
=(N!)^{-1}\sum_{\calp}(\pm)^{\calp}\delta(R_0-\calp R_*)$. The
contribution of all the paths will be of the same sign; positive
if $\rho_F (R_0, R_* ; 0) > 0$, negative otherwise. In particular,
on the diagonal, all contributions will be positive. The
``bosonic'' path integral formulation can be applied to fermion
path integrals; all that is required is to take a subset of the
bosonic paths. In principle, there exists a way to solve the
``sign problem''! We shall see that it is important to allow long,
even permutations. Macroscopic even permutations are directly
related to Fermi liquid behavior and to Bose condensation of pairs
of fermions.

The remaining problem is that the unknown density matrix appears both
on the left-hand side and on the right-hand side of 
Eq.~(\ref{RPI}) since it is used to define node-avoiding paths. To
apply the formula directly, we would somehow have to
self-consistently determine the density matrix. In practice what
we need to do is make an {\it ansatz}, which we call $\rho_T$, for
the {\it nodes} of the density matrix needed for the restriction
and use PIMC to solve the Bloch equation inside the nodal cells.
What comes out, $\rh (R', R ; \beta)$ is a solution to the Bloch
equation inside the trial nodal cells, which obeys the correct
initial conditions. It is not an exact solution to the Bloch
equation (unless the nodes of $\rho_T$ are correct) because it has
possible gradient discontinuities at the trial nodal surfaces.

The only uncontrolled approximation in the restricted path
integral method is the restriction, the rule by which we allow
paths. Clearly the success of the method hinges on the choice of
this restriction. The situation is not very different from that of
classical Monte Carlo or molecular dynamics simulations. Even if
some of the quantitative details are inaccurate, if we can
characterize the nodal restriction sufficiently well, the
simulations will be useful in understanding strongly-interacting
fermion systems and are often the most accurate computational
method available.

For the moment, let us consider the reference point $R_*$ and the
inverse temperature $t$ as fixed parameters. Then the nodes have
dimension $3N-1$ since a single equation specifies whether $R_t$
is on a node. One property that holds true in general, is that
when two fermions have both the same spin and the same spatial
coordinates, all the wave functions and hence the density matrix
must vanish. Hence, for any pair of fermions with the same spin,
the hyperplane defined by the equation: $\bfr_i = \bfr_j$ is on
the node. Since these are three equations (in three dimensions)
the ``coincident hyper-planes'' have dimensionality $3N-3$. The
coincident planes are fixed hyper-points lying on the nodal
surfaces which have a dimensionality two larger. For quantum
mechanics in one dimension, the coincident points exhaust the
nodal surfaces so that one knows the exact restriction.  For
fermions in two or three dimensions, symmetry is not sufficient to
determine the position of the nodes. Their position depends in a
non-trivial way on the potential.

Numerical investigations~\cite{ceperley91} have found the free
particle nodes (for spinless fermions) divide the path space into
a single positive region and a single negative region (except in
one spatial dimension where the nodes divide the phase space into
$N!$ regions.) This means that the restricted path partition
function includes contributions from all $N!/2$ even permutations.

\section{WINDING NUMBERS, EXCHANGE AND MOMENTUM DISTRIBUTION}

A path is a mapping from an imaginary time loop (a circle) to the
configuration space $\mathbb{R}^{3N}$. In a system of boltzmannons
(non-exchanging particles) without periodic boundary conditions,
all paths can be contracted to a point, so there is no interesting
topological structure.  But paths of identical particles in
periodic boundary conditions are can have non-trivial topologies
of their paths. Paths that wind around the periodic boundary
conditions cannot be contracted to a point. The superfluid
fraction (fraction of the mass decoupled from moving walls) in the
$x$ direction is given by
\begin{equation}
\frac{\rho_s}{\rho} = \frac{\langle W_x^2\rangle}{\hbar^2M_t\beta
},
\label{eq:winding}
\end{equation}
where $W_x=\sum_i m_i (x_i-x_{P_i})$ is the total mass-weighted
winding number in the $x$ direction. The twist free energy,
defined as the difference in free energy between periodic and
antiperiodic boundary conditions, may also be determined from the
winding of paths. Winding only occurs because of the occurrence of
long permutation cycles. The superfluid density and twist free
energy can both be used to identify Bose condensates, and both of
bosons and pairs of fermions.

A topological classification of paths is to label them by a {\em
product} of the permutation structure and the winding structure.
For example, a topological identification of a path could be that
it contains ``three 1-cycles with winding (0,0,0), two 1-cycles
with winding (0,1,0), and one 7-cycle with winding (3,2,-1).''
Topological estimators are important because they are not
concerned with the details of the paths.

Aside from the winding numbers, the permutations themselves have
physical meaning. The free energy cost to ``tag'' an atom, for
example the chemical potential of an isotopic substitution, is
\begin{equation}
\mu_\mathrm{tag} = -k_B T \log\langle n_1/N \rangle,
\end{equation}
where $n_1$ is the fraction of monomers, or 1-cycles.

The momentum distribution is the Fourier transform of the single
particle density matrix. In a translationally invariant system:
\begin{equation}
n_{\bfk} = \frac{1}{8 \pi^3 } \int d\bfr  e^{- i \bfk \cdot \bfr }
n (\bfr)
\end{equation}
where the single particle density matrix is:
\begin{equation}
n(\bfr) = \frac{1}{Z} \int dR \rho ( \bfr_1 +\bfr , \bfr_2, \ldots
, \bfr_n, \bfr_1, \bfr_2, \ldots , \bfr_N ; \beta).
\end{equation}
To get an observable in momentum space, we cannot do the
simulation entirely in the position representation. We must allow
one of the atoms to have free ends. Bose condensation maps into
the property that the two free ends will separate by a macroscopic
distance so that $n(r)$ goes to a constant at large $r$, the
fraction in the zero momentum state.

For fermions at zero temperature, the momentum distribution has a
discontinuity at the Fermi wavevector $\bfk_F$. As a consequence
the single particle density matrix must decay at large distance as
$\cos(k_F r) r^{-2}$. We can get such long-range behavior only if
there are macroscopic exchanges since a one-cycle end-to-end
distribution will decay as $\exp(-r^2/(4\lambda \beta))$. Hence
the existence of any kind of non-analytic behavior (a
discontinuity in $n_{\bfk}$ or in any of its derivatives) implies
that the restricted paths have important macroscopic permutation
cycles. Calculation of the fermion single particle density matrix
requires simulations have both even and odd permutations and hence
minus signs. It is uncertain whether the algebraic decay of $n(r)$
results from cancellation between long even and odd chains, or
whether the restriction of the paths changes the end-to-end
distribution from the delocalized distribution characteristic of
Bose-condensed systems.

Restricted paths for fermions have additional features. In the
derivation of the winding number formula for the bosonic superfluid
density, one integrates the momentum-momentum correlation function
along the path. For the most general restricted paths, such a
correlation function can only be calculated at $\beta$ and
$\beta/2$. For this reason the use of the winding number as a
measure of the superfluid density of a fermion system can give the
wrong result. If the nodal restriction is time-independent then it
can be shown that the winding number formula is a correct way of
calculating superfluid response. Time independence means that the
trial fermion density matrix factors into a product of two
wavefunctions; $\rho_T (R', R ; \beta)= \Psi_T^* (R')\Psi_T (R)$.
Such a factorization will occur in any system at a sufficiently
low temperature if the ground state is non-degenerate.

\subsection{Fermionic pairing with restricted paths}\label{sec:rpimcpairing}

We now consider the relation between the nodal surface and a phase
transition to a state with paired fermions. Previous restricted
path calculations have used the free particle density matrix for
the nodal restriction. As we show below those nodes forbid such a
transition. However, with the correct nodal surface, restricted
paths will exhibit fermionic pairing.

One clue to the role of the nodal surface
comes from considering the free Fermi gas and
the phenomena of Cooper instability. Unlike bosons, free fermions
do not undergo a phase transition. Rather, fermions undergo a
gradual process of forming a Fermi sea. In terms of restricted
paths, the free Fermi nodal surfaces prevent the type of
percolative permutation transition experienced by bosons.
Cooper~\cite{cooper56} showed that the Fermi surface is unstable
to an attractive potential and that the ground state is a
superconductor.  This state is non-perturbative, there is no
gradual transition from a Fermi liquid to a superconductor.

Free fermion nodes are rather unusual.  For two types of fermions,
``$a$'' and ``$b$,'' (in the absence of magnetic field up and down
spin particles can be considered as two separate species) the
non-interacting density matrix is a product,
\begin{equation}
\rho(R_a,R_b) = \rho^{free}(R_a)\rho^{free}(R_b),
\end{equation}
The full nodal surface is the product of the nodal surfaces of
each species alone. Because of this property a paired state is not
possible. Consider the simplest case of two ``$a$'' fermions and two
``$b$'' fermions. The full permutation space consists of 4 possible
exchanges: $I$, $P_a$, $P_b$ and $P_{ab}$ where $P_a$ is a pair
exchange of the two ``$a$'' fermions etc. The first and last
permutations are even exchanges allowed in a restricted path
simulation. We expect that if we form a tightly bound pair it will
act as a boson and that the exchange $P_{ab}$ will be allowed.
However if we try to construct such a path, the only way the
double exchange can be allowed is for both the ``$a$'' and ``$b$'' density
matrices to change sign at the same value of imaginary time. Such
a process has zero measure, so will not contribute to any integral
and hence to any observable.

Suppose $a_0$ is a point on the nodal surface of
$\rho^{free}(R_a)$ and $b_0$ on $\rho^{free}(R_b)$ This point
$(a_0,b_0)$ is a saddlepoint of the full density matrix. Let us
parameterize the region around the saddlepoint with variables $x$,
$y$, and $\mathrm{s}$, where $x$ is the coordinate perpendicular
to the node of $\rho^{free}(R_a)$, $y$ is the coordinate
perpendicular to the node of $\rho^{free}(R_b)$ and $\mathbf{s}$
are the remaining coordinates.  Then:
\begin{equation}
\rho(x,y,\mathbf{s}) = constant(\mathbf{s}) xy.
\end{equation}
Although two positive (negative) regions touch, paths are unable
to pass between them due to the nodal restriction. Any small
perturbation $\sigma$ will in general be a function of the
coordinates $\mathbf{s}$, so the density matrix in the presence of
a small interaction takes the form
\begin{equation}
\rho(x,y,\mathbf{s}) = xy + \sigma(\mathbf{s}).
\end{equation}
Any small perturbation opens holes along the singularity. The
perturbation shifts the node and saddle point so that they no
longer coincide. In particular, it is possible to prove, for the 4
interacting fermions, that one can always construct an 
exchange path $P_{ab}$ which is node avoiding.

The nodes of interacting system of two fermion types
allow significantly more permutations of restricted paths than
free particle nodes allow.  Since the free particle density
matrix is a product, the node-avoid paths must be those with
even permutations in each species.  For $N$ fermions of type ``$a$''
and $N$ fermions of type ``$b$'', free particle nodes allow
$(N!/2)^2$ permutations.  With interactions, all even permutations
may be allowed, numbering $(2N!)/2$.  Taking the classical ``entropy''
of the permutation to be $S=k\log N_\calp$, the increase in
entropy due to the extra permutations is $2Nk\log 2$.  
Although this estimate omits effects of correlation, it
does suggest that the additional permutations allowed by the nodes
of an interacting system may be enough to drive a phase transition,
much as the appearance of long permutations drive the lambda transition
in Bose systems.

\begin{figure}
\centerline{\includegraphics[width=3in]{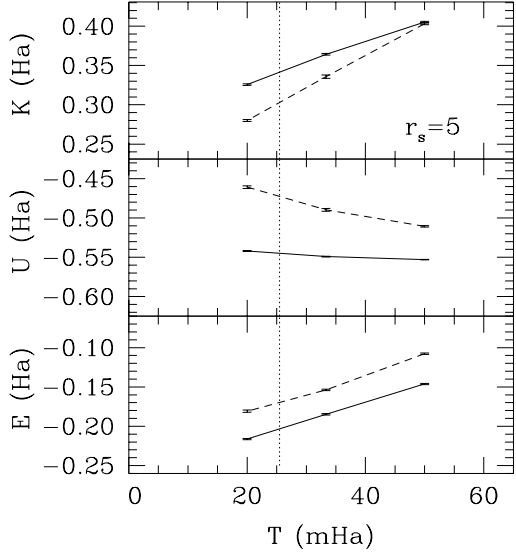}}
\caption{Effect of choice of nodal surface on the energy of a 
system of 19 electron-hole pairs at $r_s=5$. Kinetic, potential,
and total energies are shown for free particle nodes (dashed lines)
and paired Gaussian nodes (solid lines).}
\label{fig:enode}
\end{figure}

For strong pairing interactions, the nodes are quite different
from free fermion nodes.  For the case of excitons, for instance,
the wavefunction should be antisymmetric under exchange of either
``$a$'' particles or ``$b$'' particles.  A Slater determinant of paired
orbitals (such as a BCS wavefunction) has such properties, and is
a reasonable candidate for ground state nodes:
\begin{equation}
\rho(R) = \det | \phi(a_i-b_j)|,
\end{equation}
In our application to excitons discussed below we use these nodal
surfaces and denote them paired fermion nodes.
Gilgien~\cite{gilgien97} found that a Gaussian pairing function
$\phi (r) = \exp( -c r^2)$ gave lower energies than free-particle
nodes in ground state fixed node calculations of unpolarized
excitons. If these nodes describe a well-formed pair, such as in a
dilute exciton gas at very low temperature, the only way to change
sign is to exchange a pair of particles between excitons. In this
case, the nodal surfaces are localized along exciton collisions.
We have used these ground state Gaussian paired nodal surfaces in
our calculations, taking the value $c=0.3$, based on the results
of Gilgien's calculations.  In Figure~\ref{fig:enode} we compare the
energy calculated using free particle nodes and paired Gaussian
nodes.  The lower total energy of the system Gaussian nodes supports
their use for the electron-hole system near the density $r_s=5$.

RPIMC simulations are then used to determine properties of the
paired state. Since these nodes do not depend on imaginary time,
we can use the winding number formula, Eq. \ref{eq:winding}, to calculate
the superfluid density. In the case of two different species of
fermions (such as particles and holes) one can define two
different responses, the response to moving boundaries (that
couples to the mass) and the response to a magnetic field (that
couples to the charge). One can also calculate the pair condensate
fraction by cutting open both an ``$a$'' path and a ``$b$'' path. In a
Bose-condensed state, the two ends will become delocalized.
However, if the path of only the ``$a$'' particle is cut the two ends
should remain bound together.

\section{APPLICATIONS}

Bosonic calculations of superfluid $^4$He are reviewed in 
Ref.~\cite{ceperley95}. The first application of RPIMC was to liquid
$^3$He~\cite{ceperley92} using the semi-empirical Aziz potential.
Using free-particles nodes, these calculations show reasonable
agreement with experiment and the importance of using
time-dependent nodes and permutations. The interface between
superfluid $^4$He and liquid $^3$He was also
simulated~\cite{ceperley95}. The properties of an electron-proton
hydrogenic plasma  were calculated by Pierleoni {\it et al.}
(1994) and Magro {\it et al.}~\cite{pierleoni94}. From 32 to 64
electrons and an equal number of protons were put into a periodic
box, interacting with the Coulomb potential using Ewald summation.
Both the electrons and protons were fully quantum particles. A
time step of roughly $10^6$K was found adequate. Thus a simulation
at a temperatures of 4000K required 250 time slices. Good
agreement with other theoretical approaches was found at high
density and at high temperature in the plasma phase. At
temperatures below 10,000K the spontaneous formation of $H_2$
molecules, was observed, as evidenced by a strong peak in the
proton-proton correlation function at a distance of 1\AA. Militzer
will discuss the current status of those calculations in his
contribution to these proceedings. Applications of path integral
methods to two-component plasma with equal masses is discussed
below. PIMC has also been used to calculate the melting
temperature of the quantum OCP showing a higher maximum melting
temperature than had been previously thought~\cite{jones96}.

These various simulations demonstrate the power of the method.
Once the RPIMC method is programmed, one can rather directly do
highly accurate simulations of experimentally relevant systems
without tedious construction of basis functions and trial
functions; the type of systems which can be treated are much more
complex than can be handled with other QMC techniques,
and potentially more accurate than mean field approximations.

\section{ELECTRON-HOLE PLASMAS}

In the remainder of this contribution we want to explain some
computer experiments to explore the idea of fermionic pairing
leading to Bose condensation, in particular the picture one is led
to in restricted paths. Although BCS
pairing~\cite{cooper56,bardeen57a,bardeen57b} is a convenient way
to explore how the weak pairing of fermions leads to
superconductivity, the energy and lengths scales involved make it
difficult to explore the weak fermion pairing with direct quantum
simulations. The simplest and most suitable system to study
fermionic pairing in a plasma system is the two component plasma:
a system with two species of oppositely charged fermions. The
maximum transition temperature will occur if the fermions are
equally massive and spin-polarized.

One realization of the two-component plasma is in the realm of
low-temperature semiconductor physics \cite{wolfe95}. Conduction
electrons and holes in semiconductors interact with Coulomb force
and can have very similar effective masses.  Temperatures can be
lowered far enough that the thermal wavelength exceeds the
interparticle spacing so that it is a degenerate quantum plasma
with some evidence of Bose condensation~\cite{snoke90,lin93}.
However, the short lifetimes of the excitations due to
recombination, and interaction of the plasma with the
semiconductor lattice and exciting laser pulse complicate the
analysis. Recent work by O'Hara~\cite{ohara99} suggests that
experimental exciton densities may not be high enough for Bose
condensation, and the observed spectral features are due instead
to dynamic effects.

\begin{figure}[t]
\sidebyside{\includegraphics[width=0.9\linewidth]{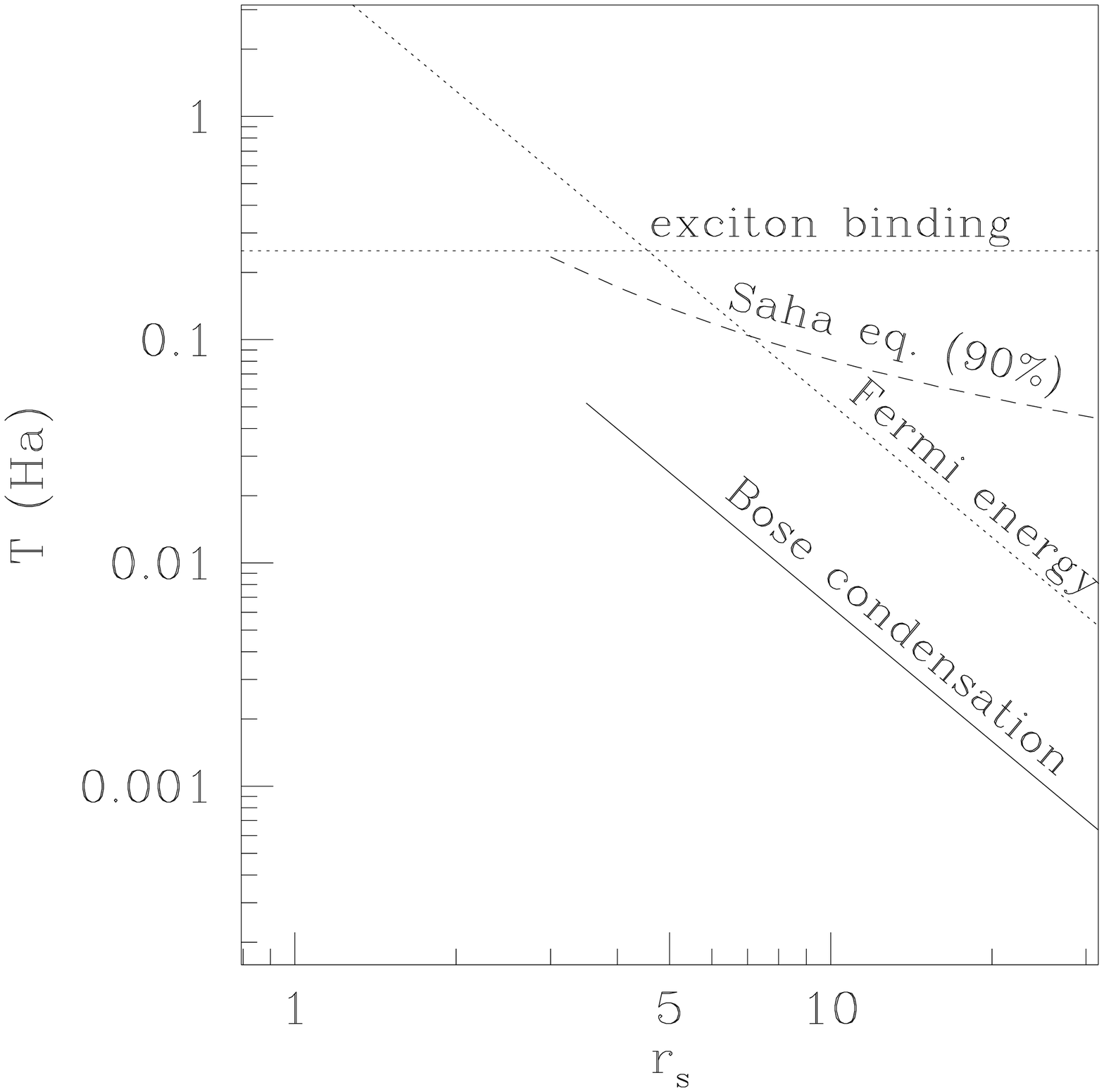}}
           {\includegraphics[width=0.9\linewidth]{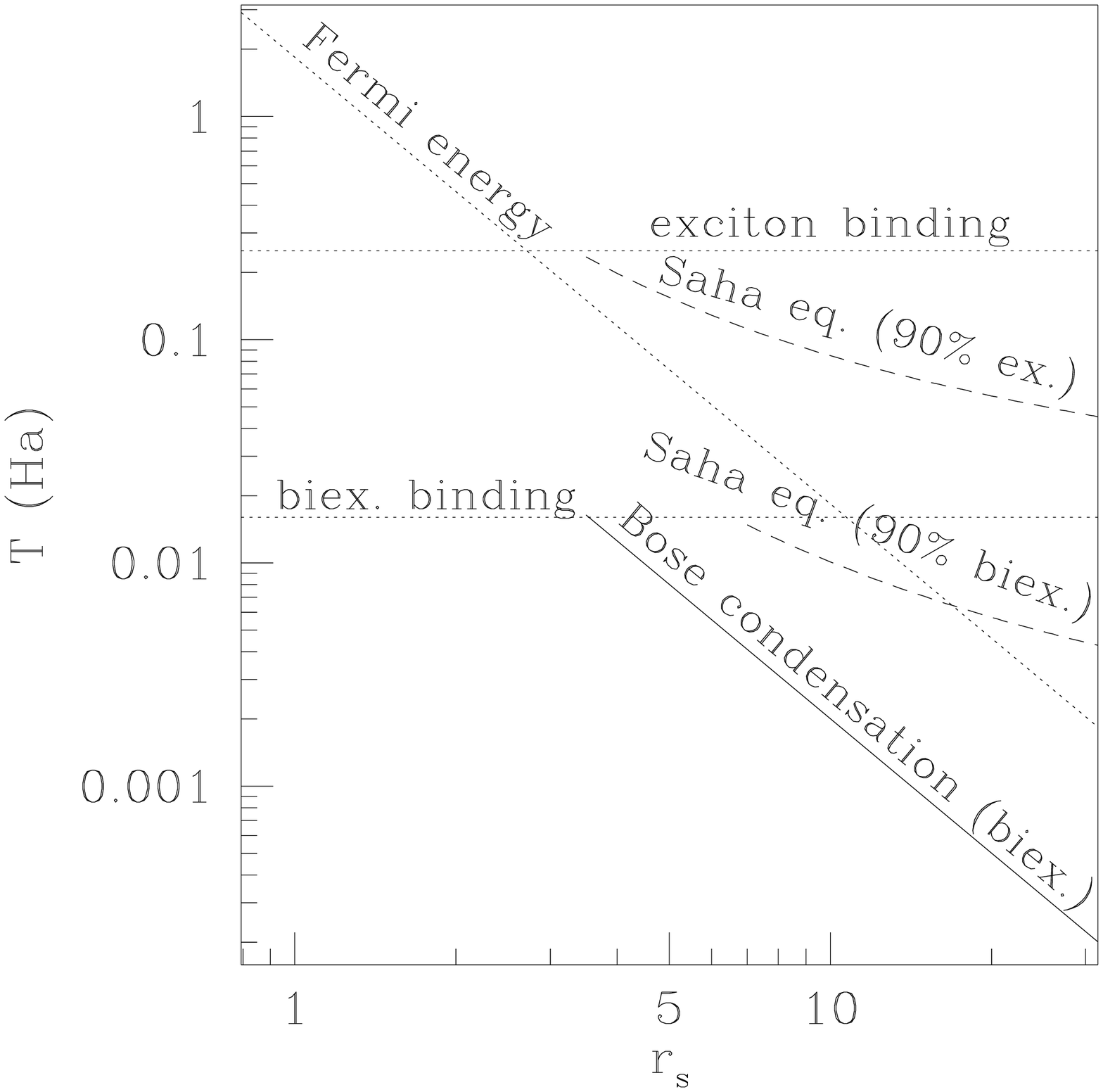}}
\caption{The phase diagram of the equal-mass two-component plasma,
for (a) spin-polarized particles and (b) unpolarized particles.
Horizontal line is the exciton binding energy, and the solid diagonal lines
are the ideal Bose transition temperatures, $T_c^0$
for (a) spin polarized excitons and (b) biexictons. The curved
dashed lines are predictions of the Saha equation for the appearance
of excitons and biexcitons.}
\label{sketch}
\end{figure}

A sketch of the phase diagram of an electron and hole plasma is
shown in Figure \ref{sketch} for polarized and un-polarized
excitons.

A natural energy unit for this problem is the Hartree (Ha) $m_e
e^4/\epsilon^2\hbar^2$ and length unit is the Bohr radius,
$\hbar^2\epsilon/m_e e^2$.  In these units the exciton binding
energy is $E_x=0.25$~Ha and the exciton radius is $a_x=2a_0$. The
dimensionless quantity $r_s$, is defined so that the density of
electron-hole pairs is $(\frac{4}{3}\pi r_s^3a_0^3)^{-1}$. At
temperatures below $E_x \approx 0.25$ Ha, electron hole pairing
can lead to an excitonic regime.  However, the formation of
excitons is inhibited either by the entropy of unbound states at
low densities and both Coulombic and exchange interactions between
at high density.  In the limit of high density $r_s < 1$ the
exciton gas will approach two ideal Fermi liquids of opposite
charges because the kinetic energy will dominate over the
potential energy.

Bose condensation of excitons is expected at a sufficiently low
temperature.  Note that there can be a qualitative change in the
transition: at low densities excitons form above the transition
temperature and then Bose condense at lower temperatures, while at
high densities pairs may not be bound, but form only because of
condensation, as in BCS theory, as discussed by
Randeria~\cite{randeria95}.  One open question is whether this
transition is continuous as density is increased~\cite{leggett80}
or whether there are further transitions such as the ``plasma
phase transition'' speculated to exist in hydrogen.

In a spin-polarized system, all excitons are identical, so the
quantum degeneracy is higher than in an unpolarized system and so
should have a higher transition temperature (see
Figure \ref{sketch}) (remember unpolarized excitons have four
possible spin states.) Also, excitons are not stable in the
unpolarized system, but they pair up and form biexcitons,
consisting of two electrons and two holes, with one spin-up and
one spin-down particle of each species.  Although the biexcitons
are also interesting candidates for Bose condensation, the
transition temperature will occur at an order of magnitude lower
temperatures. In this work we consider equal numbers of electrons
and holes of the same mass interacting with a Coulomb potential in
a cubic periodic box. We have done calculations with both
spin-polarized and unpolarized systems, but the pairing
calculations (using the paired nodal surfaces discussed earlier)
are only of the polarized systems.

At low temperature and low density, the spin-polarized electron-hole
system forms a dilute Bose gas, made up of spin-polarized excitons.
In the absence of exciton-exciton interactions, the transition
temperature for Bose-Einstein condensation (BEC) is
\begin{equation}
T_c^0 = 1.275\frac{\hbar^2}{m k_B a_0^2 r_s^2}
\end{equation}
where $m=2$ is the exciton mass.  Repulsive interactions increase
$T_c$ in a dilute Bose gas~\cite{grueter97}. Using the
exciton-exciton scattering length $a_s = 1.5~a_x = 3.0~a_0$ to
compare with the hard sphere calculations of Grueter {\em et
al.}~\cite{grueter97}, we expect an enhancement in $T_c$ of about
5\% above the non-interacting transition temperature, $T_c^0$.

We begin by presenting our simulation evidence of biexcitons in
unpolarized system. We then discuss the BEC transition in
spin-polarized systems.

\subsection{Evidence of biexcitons in unpolarized systems}

\begin{figure}[t]
\centerline{\includegraphics[width=.5\linewidth]{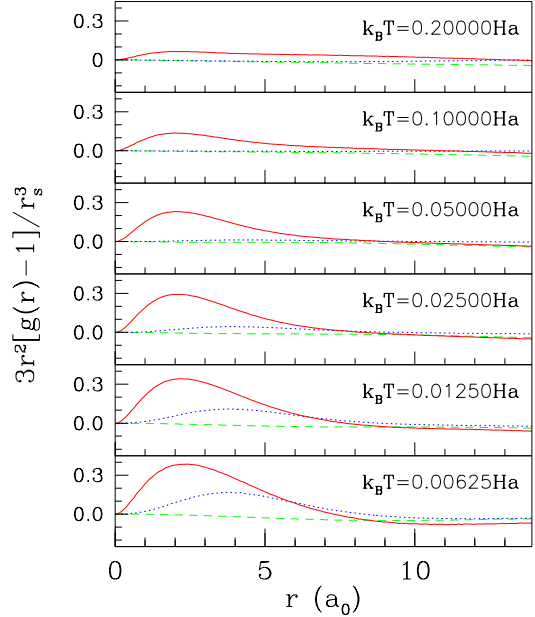}}
\caption{Relative pair density functions for the spin-unpolarized
electron-hole system at $r_s=10$, showing 
formation of excitons and biexcitons. Correlation functions
$3r^2[g(r)-1]/r_s^3$ are shown for opposite species (solid lines), same
species with opposite spin (dotted lines), and same species with
same spin (dashed lines).  Normalization is chosen so that area
under the curve represents the change in density relative to a
uniform distribution.  At $T=0.05$~Ha the area under the opposite
species curve is approximately one, indicating excitons. At
$T=0.00625$~Ha the area under the opposite species curve is
approximately two, and the area under the same species, opposite
spin curve is approximately one, indicating biexcitons.}
\label{fig:biexgr}
\end{figure}

In Figure~\ref{fig:biexgr} we show evidence of exciton and
biexciton formation in a system of fourteen spin-unpolarized
electron-hole pairs at a density $r_s=10$ computed using free
particle nodal surfaces. The figure shows the change in the pair
correlation functions, normalized so that the area under the curve
represents the radial change in density relative to a uniform
distribution. Electrons and holes attract each other, and the
location of the peak near $r=2 a_0$ agrees very well with the
exciton radius, $a_x=2a_0$. The area under the curve at
$T=0.05$~Ha is approximately one, so we interpret this as an 
excitonic regime.

Also shown in Figure~\ref{fig:biexgr} is the deviation of the pair
correlation function from a uniform distribution between particles
of the same species.  The dotted lines are same species with
opposite spin, and the dashed lines are same species with the same
spin. Although the relative density for same species with
like spin always dips negative, indicating a depletion, particles
of the same species with opposite spin have some attraction.  At
$T=0.00625$~Ha the area under the same species, opposite spin
curve is approximately one, while the area under the opposite
species curve has increased to two. This means that a particle is
likely to have three other particles near it: one particle of the
same species with opposite spin, and two particles of the other
species (one in each spin state).  We interpret this as formation
of biexcitons.

To summarize, for a system of spin-unpolarized electron-holes at a
density of $r_s = 10$, we find excitonic formation beginning below
$T\approx0.2$~Ha $=0.8 E_x$, with an excitonic regime formed by
$T\approx 0.05$~Ha $=0.2 E_x$.  Below this temperature, excitons
begin to pair together, and by $T\approx 0.00625$~Ha $=0.025 E_x$
we see  biexcitons form.

\subsection{Permuting excitons and the superfluid transition}

As discussed previously, Bose condensation of excitons is expected
to be seen as long cycles of permuting electrons.   The excitons
should appear as electron and hole paths propagating side by side
in imaginary time.  One way to look for Bose condensation is to
look for such a feature in a graphical representation of the
paths.

\begin{figure}[t]
\sidebyside{\includegraphics[width=0.8\linewidth]{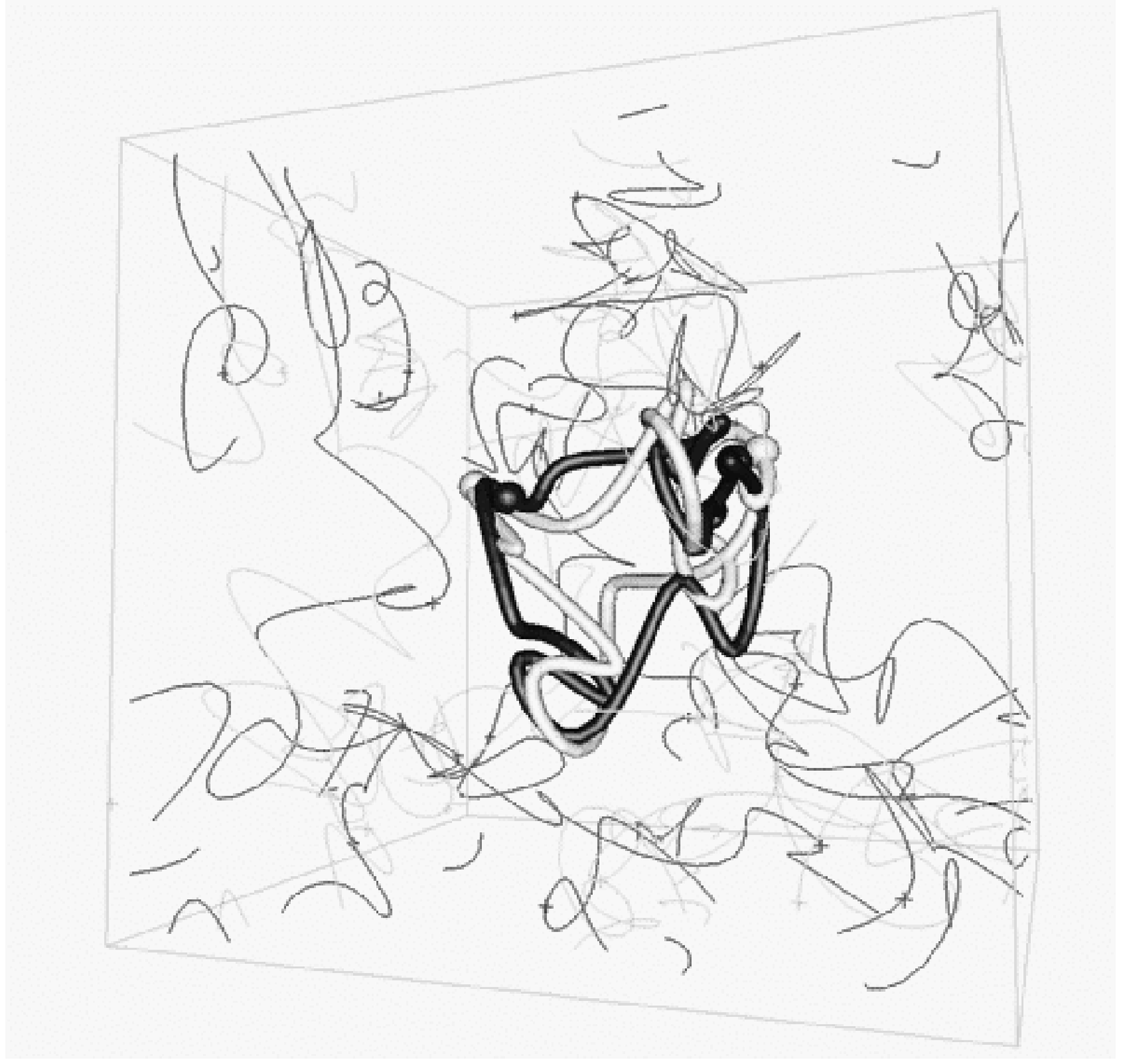}}
           {\includegraphics[width=0.8\linewidth]{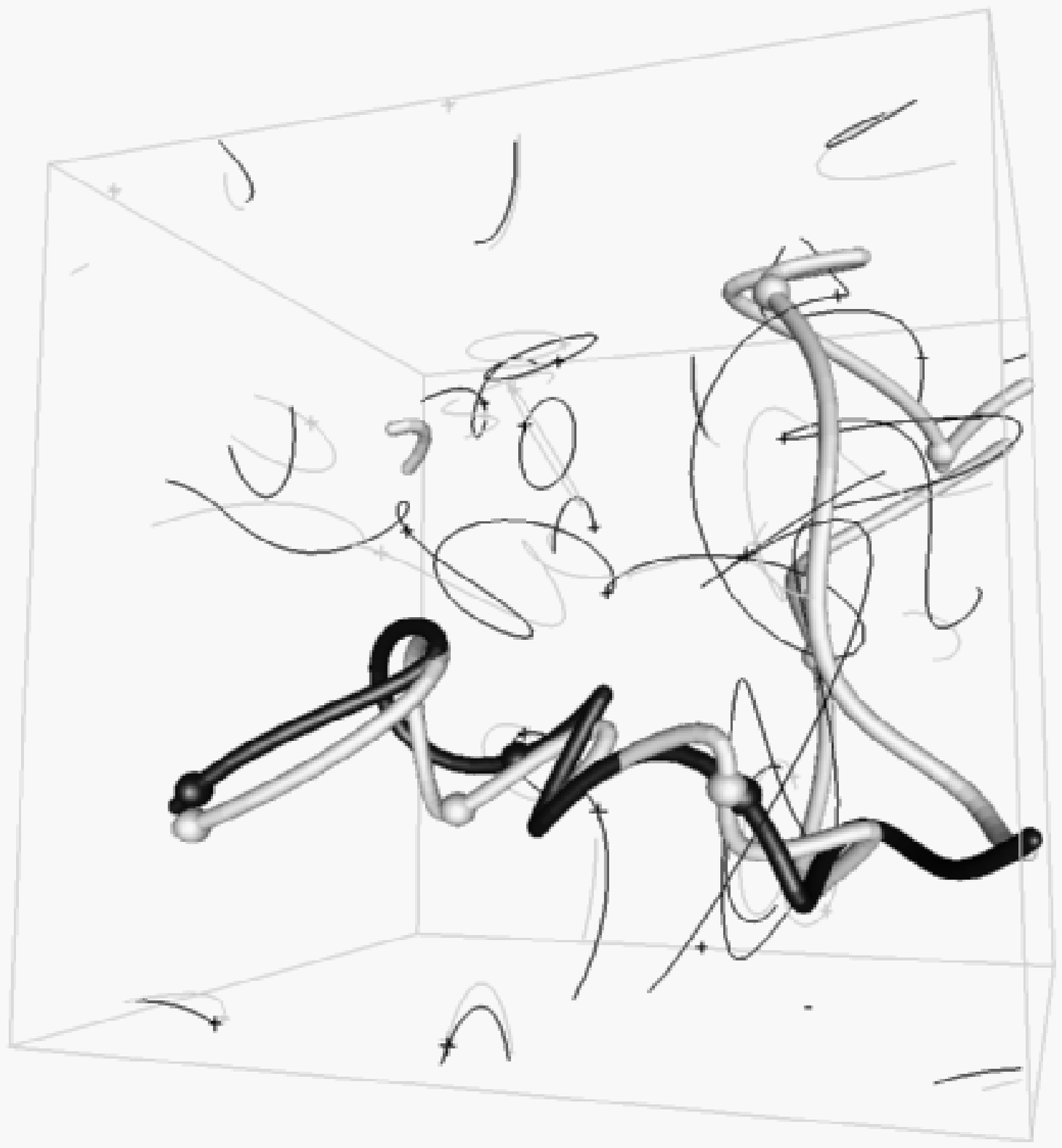}}
\caption{A ``snapshot'' taken from a PIMC simulation of 19
spin-polarized electron-hole pairs at $r_s=6$ and $T=0.0125$~Ha.
To enhance visual clarity, Fourier smoothing has been used to
eliminate the imaginary- time high frequency noise. The
dark-colored paths represent electrons and the light-colored paths
represent holes. The thick paths in the panels show: (a) two
permuting excitons, (b) a winding electron-hole pair. The thin
paths represent all the other particles in the simulation. Paired
nodal surfaces were used for these calculations. }
\label{fig:permpict}
\end{figure}

In Figure \ref{fig:permpict} is a ``snapshot'' from a simulation
of 19 electron-hole pairs, one panel high-lighting a pair of
permuting excitons, the other 3 permuting excitons winding around
the boundaries. What is interesting in this picture is the
complexity of the fermion pairing.  Close examination shows that
electrons and hole are almost always paired up. However one does
not simply have pairing between two like membered exchanges of the
same length. Rather there is an intertwined network of coupled
electron and hole exchanges. This is reflected in the mass and
charged coupled winding statistics.

\begin{figure}[t]
\centerline{\includegraphics[width=.5\linewidth]{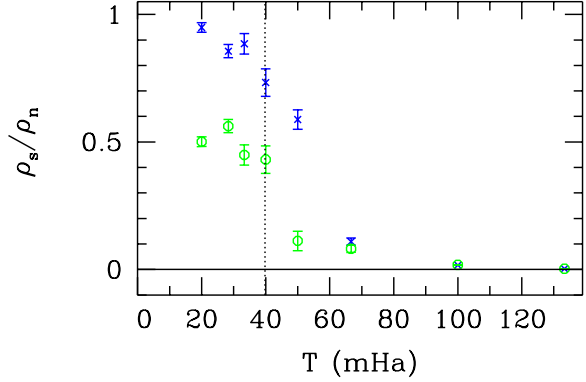}}
\caption{Superfluid density: mass coupled
superfluid density, charge-coupled superfluid density, for
electrons and holes at $r_s=4$. }
\label{fig:sfd}
\end{figure}

Shown in Figure \ref{fig:sfd} are the mean squared winding numbers both for
mass and charge. A superfluid is defined as a system that
decouples from a boost in the boundaries: the appropriate winding
number to compute is to add the windings for the electrons and
holes before one squares. This number shows a rapid
increase near the peak in the specific heat and rapidly approaches
unity.  An estimate of $T_c$ can be determined by a finite size
scaling study of the superfluid density.

The second curve shows the response of the system to a weak
imposed magnetic field. A superconductor will exclude magnetic
fields: the Meisner effect. A magnetic field couples to the
charge, so the appropriate winding number is the electron winding
minus the hole winding. For tightly bound excitons, the electron
winding would always equal the hole winding, since one has a neutral
object which does not response to the applied field. We find a
partial response, substantial screening does occur. Further
studies need to be made to understand if this curve scales to zero
in the thermodynamic limit. In a system of unpolarized particles,
one would have additional spin response functions.

\begin{figure}[t]
\sidebysidenl{\includegraphics[width=.8\linewidth]{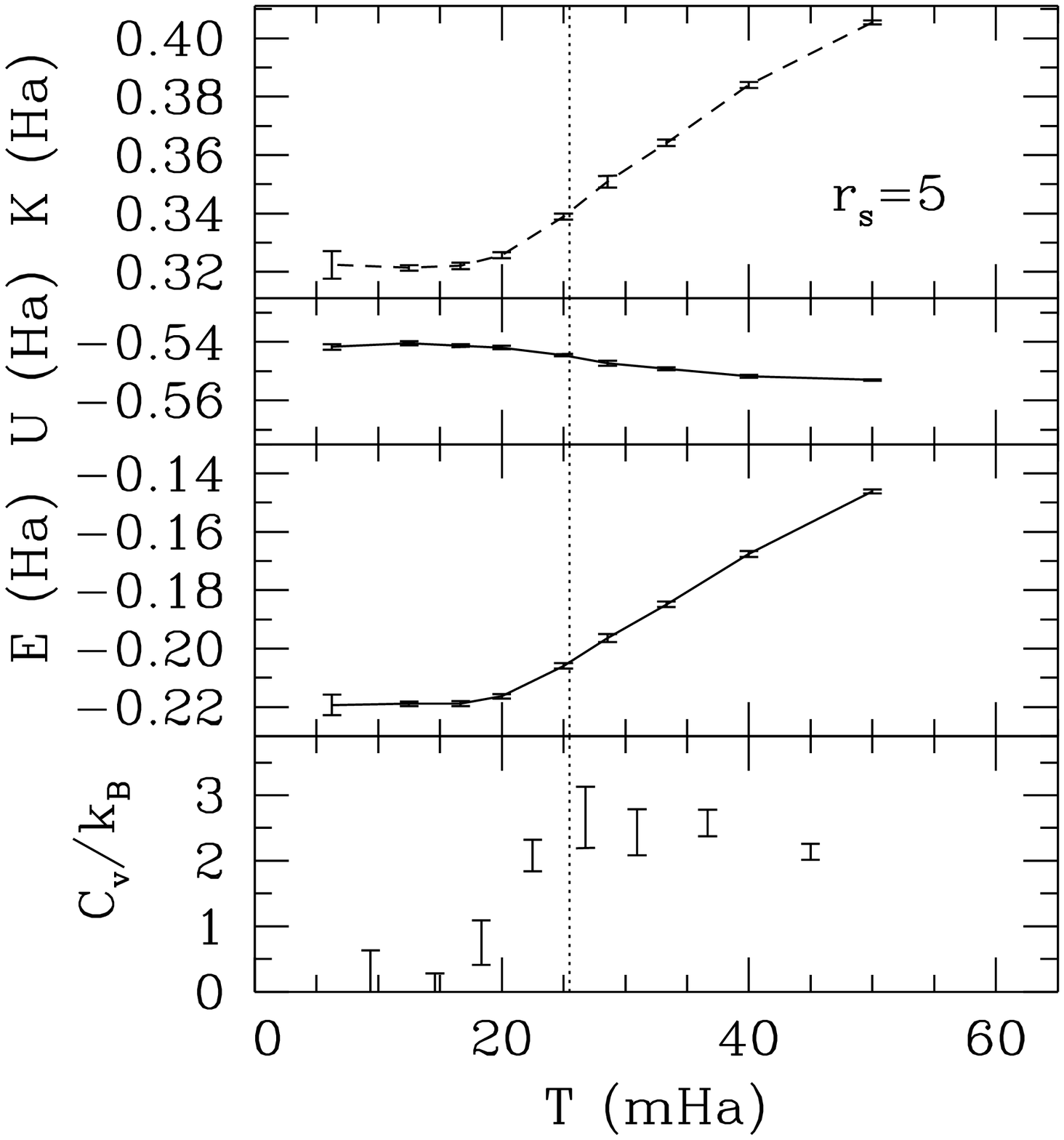}}
             {\includegraphics[width=.8\linewidth]{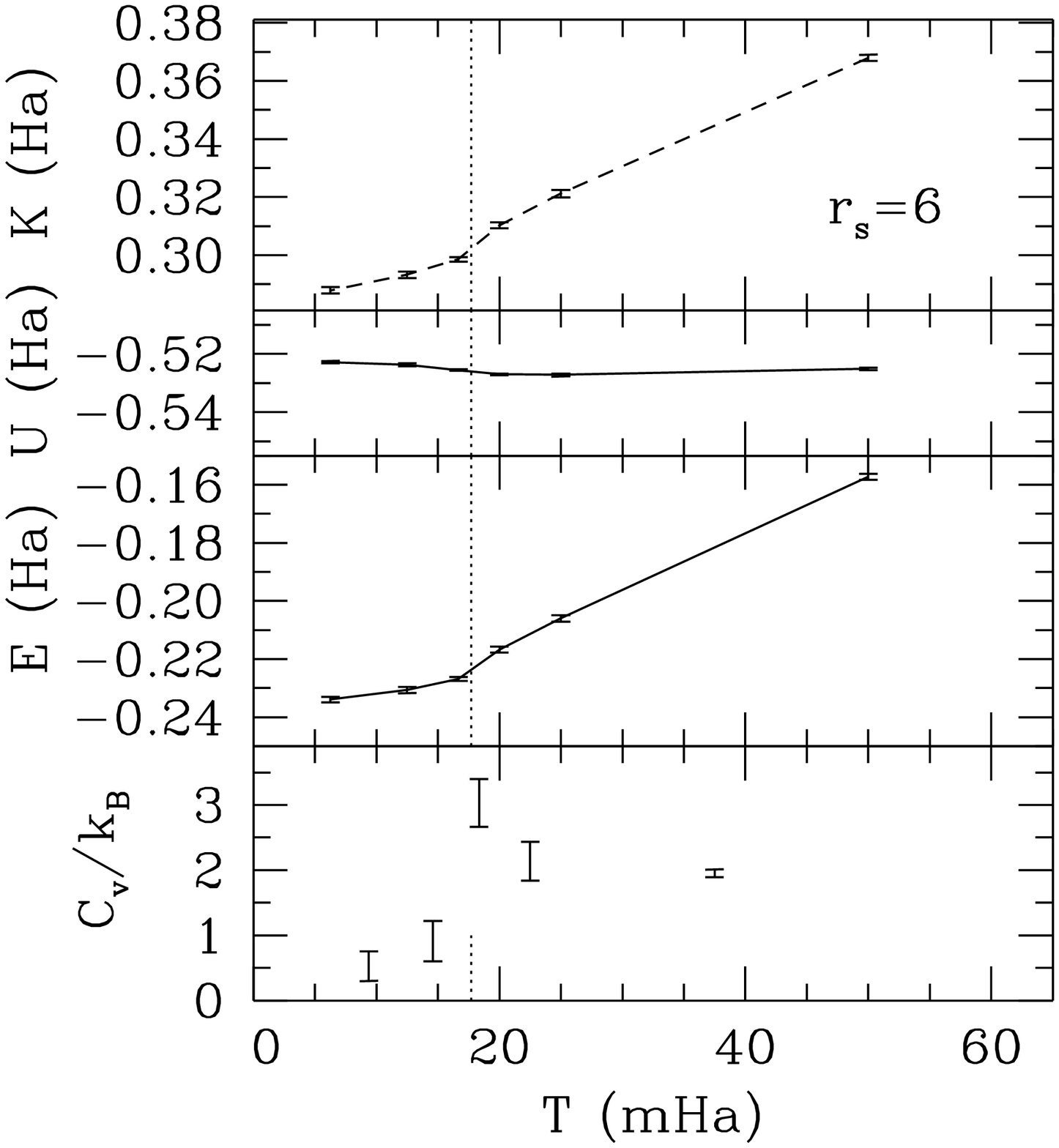}}
\caption{Kinetic, potential and total energy, and specific heat for 19
spin-polarized excitons as a function of temperature, at 
densities $r_s=5$ and $r_s==6$. Vertical dashed lines are the ideal gas
transition temperature for each density.  Densities are:
$r_s=6$ (left) and  $r_s=5$ (right).}
\label{fig:lambda_evst}
\end{figure}

A characteristic feature of Bose condensation is the peak in the
specific heat at the superfluid transition. Not only is the
feature prominent in both weakly and strongly interacting bosons,
even superconducting transitions exhibit a similar peak.

We plot the specific heat as estimated from numerical derivatives
of our total energy values in Figure~\ref{fig:lambda_evst}.  For
$r_s=6$ we see a clear peak near the transition line, and similar
behavior for $r_s=5$. For higher density of $r_s=4$ the data is
much too noisy show any real features but we do see signs of the
lambda transition. Since the specific heat is a derivative of the
energy, statistical errors in the energy get magnified.  The
observation of the lambda transition in the specific heat is
computationally demanding and not a particularly good indicator of
the phase transition.

In Figure~\ref{fig:lambda_evst} we plot the kinetic, potential,
and total energies as a function of temperature for densities
$r_s=6,5$ and 4.  For reference, we show the transition
temperature $T_c^0$ for a non-interaction system as a vertical
dotted line in each of the figures.  We see qualitatively
different behavior in the energy above and below $T_c^0$. The
kinetic energy changes slope, from nearly constant below $T_c^0$
to a slope in the range $2 k_B$ to $3 k_B$ above $T_c^0$.  The
potential energy becomes slightly less negative below $T_c^0$. We
interpret the different kinetic energy behaviors as changing from
a classical thermal distribution of excitonic motion (and internal
excitations of the excitons) above $T_c^0$ to a Bose condensed
state with little excess kinetic energy below $T_c^0$.

The flat behavior of the energy of low temperature is an
indication of Bose like (not Fermi) excitation spectra. Once one
is below the transition, there are few excitations.

\subsection{Energy at low temperature}
\begin{figure}[t]
\centerline{\includegraphics[width=.5\linewidth]{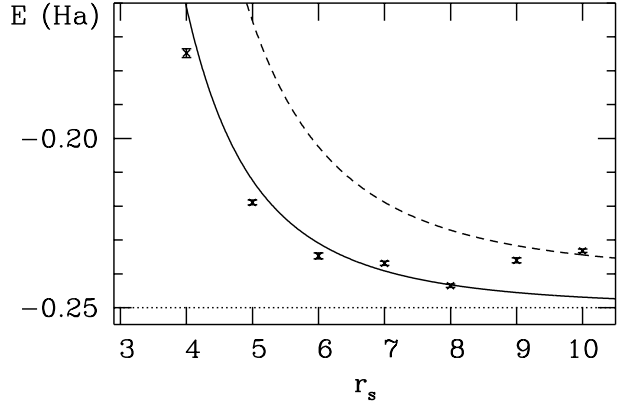}}
\caption[Condensate energy as a function of density for
spin-polarized excitons.] {Condensate energy as a function of density for
spin-polarized excitons. Solid line is Bogoliubov theory, data points
are from PIMC.}
\label{evsrsFig}
\end{figure}

At low temperature and low density, the interactions between
excitons are described by the exciton-exciton scattering length,
$a_s=3.0a_0$, as calculated in Ref.~\cite{shumway99}. We have
calculated the total energy of 19 spin-polarized electron-hole
pairs at $T=0.00625$~Ha and a range of densities,
$r_s=4,5,6,7,8,9$, and $10$ using paired nodal surfaces. We plot
the total energies in Figure~\ref{evsrsFig}.

The ground state energy of Bose condensed excitons well below the
transition temperature is given by Bogoliubov theory for a dilute
Bose gas~\cite{Abrikosov63},
\begin{equation}
\frac{E}{V} = \frac{2\pi a_s}{m} \left( \frac{N}{V} \right)^2
\left[ 1 + \frac{128}{15\sqrt\pi} a_s^{3/2}
\left(\frac{N}{V}\right)^{1/2} \right]
\approx\frac{2.25}{r_s^3}\left[1+\frac{12.2}{r_s^{3/2}}\right].
\end{equation}
$N$ is the number of bosons, and $V$ is the volume of the
(periodic) box. The last expression is for spin-polarized excitons
with a mass $m=2$ and and a scattering length of $a_s = 1.5 a_x =
3.0 a_0$, as found in Ref.~\cite{shumway99}. This assumes that
only two particle collisions are significant, $s$-wave scattering
dominates, and only interactions involving the condensate are
important. The first two assumptions hold for a low-density
system, $N/V \ll a_s^{-3}$. The last assumption is valid when the
condensate fraction is large, so that the energy contribution due
to the interactions between non-condensed particles is negligible.

Simulations with $r_s\le 8$ agree very well with the theory, (note
there are no adjustable parameters) as shown in Figure
\ref{evsrsFig}. Simulations at lower densities, have higher
energies because of thermal effects. The transition temperature
decreases with density, taking on the values $T_c^0=0.07875$ and
$T_c^0=0.06375$ at $r_s=9$ and $r_s=10$, respectively.  Few of the
excitons are condensed at these densities, hence the energies are
above the Bogoliubov ground state energy.  We estimate the energy
per pair of non-condensed excitons as the exciton binding energy,
plus $3/2 k_BT$ for the kinetic energy of the excitons, plus {\em
twice} the Bogoliubov interaction energy (the factor of
two arises from the exchange term, which is not present in the condensate)
and plot that as a dotted line in the figure.  The
agreement confirms our interpretation of the data.

\section*{ACKNOWLEDGMENTS}
Computer runs were made at the NCSA. We are supported by the
Department of Physics at UIUC and  by NSF through grants DMR
98-02373 and DGE 93-54978. Valuable assistance has been provided
by Greg Bauer, Burkhard Militzer and Mark Dewing.

\bibliographystyle{prsty}
\setlength{\itemsep}{-10pt}
\bibliography{ex_plasma}
\end{document}